\documentclass[journal,12pt,onecolumn,draftclsnofoot]{IEEEtran}

\IEEEoverridecommandlockouts

\usepackage{pgfplots}
\usepackage[bookmarks,colorlinks]{hyperref}
\usepackage[linesnumbered,ruled,lined]{algorithm2e}
\usepackage{enumitem}
\usepackage{algpseudocode}
\usetikzlibrary{shapes.multipart,intersections}
\usepackage{cite}
\usepackage{amsmath,amssymb,amsfonts,steinmetz}
\usepackage{graphicx}
\usepackage{mathrsfs}  
\usepackage{textcomp}
\usepackage{acronym}
\usepackage{xcolor}

\usepackage{tikz}
\usetikzlibrary{calc}
\makeatletter
\newcommand{\gettikzxy}[3]{%
  \tikz@scan@one@point\pgfutil@firstofone#1\relax
  \edef#2{\the\pgf@x}%
  \edef#3{\the\pgf@y}%
}
\makeatother

\usepackage[draft]{todonotes}

\usepackage{esvect}

\usepackage{times}
\usepackage{bm}
\usepackage{amsmath}
\usepackage{amssymb}
\usepackage{stmaryrd}
\usepackage{babel}
\usepackage{graphics, graphicx}
\usepackage{xcolor}
\usepackage{gensymb}
\usepackage{cite}
\usepackage{enumitem}
\usepackage{url}
\newtheorem{rem}{Remark}

\acrodef{cdf}[CDF]{coupled-dipole formalism} 
 
\begin{document}
\title{PhysFad: Physics-Based End-to-End Channel Modeling of RIS-Parametrized Environments with Adjustable Fading}

\author{Rashid Faqiri, Chlo\'e Saigre-Tardif, George C. Alexandropoulos,~\IEEEmembership{Senior~Member,~IEEE}, Nir Shlezinger,~\IEEEmembership{Member,~IEEE}, Mohammadreza F. Imani,~\IEEEmembership{Member,~IEEE}, and Philipp del Hougne
\thanks{
R. Faqiri, C. Saigr\'e-Tardif, and P. del Hougne are with Univ Rennes, CNRS, IETR - UMR 6164, F-35000, Rennes, France (e-mail: \{rashid.faqiri; chloe.saigre-tardif; philipp.del-hougne\}@univ-rennes1.fr).
}
\thanks{G. C. Alexandropoulos is with the Department of Informatics and Telecommunications,
National and Kapodistrian University of Athens,  15784 Athens, Greece (e-mails: alexandg@di.uoa.gr).
 }

 \thanks{
N. Shlezinger is with the School of ECE, Ben-Gurion University of the Negev, Beer-Sheva, Israel (e-mail: nirshl@bgu.ac.il).}
 
\thanks{
M.~F.~Imani is with the School of ECEE, Arizona State University, Tempe, AZ 85281 USA (e-mail: mohammadreza.imani@asu.edu).

R. Faqiri and C. Saigr\'e-Tardif contributed equally. \textit{(Corresponding Author: Philipp del Hougne.)}
}
\vspace{-0.95cm}
}

\maketitle
	\pagestyle{plain}
	\thispagestyle{plain}
\vspace{-0.95cm}	
\begin{abstract}
\vspace{-0.1cm}

Programmable radio environments parametrized by reconfigurable intelligent surfaces (RISs) are emerging as a new wireless communications paradigm, but currently used channel models for the design and analysis of signal-processing algorithms cannot include fading in a manner that is faithful to the underlying wave physics. 
To overcome this roadblock, we introduce a physics-based end-to-end model of RIS-parametrized wireless channels \textit{with adjustable fading} (coined \textit{PhysFad}) which is based on a first-principles coupled-dipole formalism. PhysFad naturally incorporates the notions of space and causality, dispersion (i.e., frequency selectivity) and the intertwinement of each RIS element's phase and amplitude response, as well as any arising mutual coupling effects including long-range mesoscopic correlations. PhysFad offers the to-date missing tuning knob for adjustable fading. We thoroughly characterize PhysFad and demonstrate its capabilities for a prototypical problem of RIS-enabled over-the-air channel equalization in rich-scattering wireless communications. We also share a user-friendly version of our code to help the community transition towards physics-based models with adjustable fading.
\end{abstract}

\vspace{-0.45cm}
\begin{IEEEkeywords}
\vspace{-0.1cm}
Reconfigurable intelligent surfaces, end-to-end channel modeling, fading channels, discrete dipole approximation, over-the-air equalization.
\end{IEEEkeywords}

\section{Introduction}

Wireless communication systems traditionally consider the wireless propagation environment to be an uncontrolled variable. Recently, 
a paradigm shift originated from the idea of using programmable metasurfaces as “reconfigurable intelligent surfaces” (RISs) to control the wireless environment. Precursors of this ``smart radio environment'' concept emerged in the early 2000s~\cite{sievenpiper2003two,holloway2005reflection} as well as roughly ten years later~\cite{subrt2012intelligent,Kaina_metasurfaces_2014,CQW14}. 
More recently, these concepts were introduced in the wireless communications community~\cite{Liaskos_Visionary_2018,huang2019reconfigurable,di2019smart,wu2019towards,basar_Wireless_2019} and are now envisioned to become
a pillar of future sixth Generation (6G) wireless communications~\cite{Samsung}.

The role of RIS in wireless communication systems is to shape the wireless channels. 
For operation in free space, RISs are mainly deployed together with a well-aligned wave source as part of the transmit architecture, in order to implement beamforming and information encoding without costly phased-array hardware~\cite{CQW14,dai2020reconfigurable,tang2020mimo}. For operation in quasi-free space with a blocked line-of-sight (LOS) between transmitter and receiver, RISs are mainly deployed as an alternative relaying mechanism~\cite{huang2019reconfigurable,tang2020wireless,arun2020rfocus,di2020reconfigurable}. In rich-scattering environments~\cite{alexandg_2021}, where multiple scattering yields a seemingly random superposition of reflected waves with all possible angles of arrival and polarizations, RISs are used to purposefully perturb the ``disorder'' to create a monochromatic~\cite{Kaina_metasurfaces_2014,dupre2015wave} or time-coherent polychromatic~\cite{del2016spatiotemporal,imani2021smart} focus, for signal-to-noise (SNR) enhancement or over-the-air equalization, respectively, as well as to optimize the rank of multiple-input multiple-output (MIMO) channels~\cite{del2019optimally}. Besides these use cases in ``active''  communication, RISs are also used for encoding information into existing ambient waves in ``passive'' backscatter communication~\cite{zhao2020metasurface,f2020perfect}.

The successful deployment of RISs largely depends on the development of signal processing tools~\cite{bjornson2021reconfigurable}. In the development of such tools, it is of uttermost importance to ensure that the underlying channel models are compatible with the experimental reality; otherwise, it may turn out later that the developed algorithms do not function in real life, or do not perform as well as expected. This gives rise to the need for end-to-end channel models that faithfully capture the wave physics involved in programmable wireless environments parametrized by RISs. Such models begin to emerge for the operation of RISs in free space \cite{tang2020wireless,zhao2020metasurface,danufane_PathLoss_2021,di2021communication,gradoni_EndtoEnd_2020,abeywickrama2020intelligent} (see Sec.~\ref{subsec:Free} for details). However, free space is a very simple propagation environment without any of the complicated fading effects often encountered in reality. 
In the sub-$6$~GHz regime, even office rooms give rise to substantial reverberation~\cite{Kaina_metasurfaces_2014,del2019optimally}; this rich scattering results in multipath wireless channels clearly beyond the free-space approximation.
Although reverberation is weaker at higher millimeter-wave frequencies due to stronger absorption by walls, many deployment scenarios for RIS-assisted wireless communication involve operation inside metallic scattering enclosures (e.g., vessels, trains, and planes) for which the free-space approximation is unsuitable even for millimeter waves. Thus, the benefits of RIS-parametrized wireless channels cannot be fully explored and reaped based on free-space models. At the same time, conventional models of fading channels are inherently of a \textit{statistical} nature, and thus incompatible with the \textit{deterministic} control that the RIS implements as part of the scattering environment (see Sec.~\ref{subsec:AdHoc} for details). 
An end-to-end channel model that faithfully emulates wave propagation in RIS-parametrized environments \textit{with adjustable fading} is to date missing, let alone openly accessible to the community.

In this work, we fill the above-identified gap by developing an end-to-end channel model for RIS-empowered wireless communications \textit{with adjustable fading} that fully complies with wave physics. We coin our model {\em PhysFad} and share it as an open-source software. 
PhysFad enables the community to explore the potential of RIS-parameterization of wireless channels beyond the simple free-space case. In a first use case, the input parameters of PhysFad can be chosen to represent generic wave scattering problems to output channel realizations that replace those originating from conventional random-matrix approaches. PhysFad's channel statistics can obey well-known fading models (e.g., Rician fading, see Sec.~\ref{subsec:AdFad} for details) but, unlike statistical models, the channels will correctly include RIS parametrization, frequency selectivity, causality, and mesoscopic correlations~\cite{hsu2017correlation}. In a second use case, the input  parameters of PhysFad can be judiciously chosen to study specific scenarios with specific antennas characteristics, RIS designs, and scattering environments (e.g., geometry, reflectivity, and absorption). The generality of PhysFad implies that it can also readily be leveraged in backscatter-communication problems.

The paper is structured as follows: 
In Sec.~\ref{sec:sota}, we succinctly review the state-of-the-art on channel modeling. 
In Sec.~\ref{sec:ChModel}, we introduce PhysFad's channel model. We begin by introducing the underlying coupled-dipole formalism (Sec.~\ref{subsec:ChModCDF}) and explain how it can model the basic entities in wireless communications, i.e., transceivers (Sec.~\ref{subsec:ChModTransceivers}), the wireless environment (Sec.~\ref{subsec:ChModEnvironment}), and RISs (Sec.~\ref{subsec:ChModRIS}). We combine these elements into an end-to-end channel matrix formalism (Sec.~\ref{subsec:ChModMatrix}) and illustrate the implementation of adjustable fading (Sec.~\ref{subsec:AdFad}). We provide a succinct algorithmic summary for PhysFad in Sec.~\ref{subsec:AlgSumm}.
In Sec.~\ref{sec:TDRep}, we demonstrate PhysFad's time-domain capabilities. 
In Sec.~\ref{sec:CaseStudy}, we present a case study on RIS-enabled over-the-air channel equalization which showcases PhysFad's abilities: \textit{i}) to include RIS parametrization; \textit{ii}) to include channel fading; and \textit{iii}) to yield causal time-domain responses. 
Finally, we discuss PhysFad's open-source code availability (Sec.~\ref{sec:Software}) and provide concluding remarks(Sec.~\ref{sec:Conc}).

\section{State-of-the-Art and Our Contribution}
\label{sec:sota}

A channel \textit{model} is a mathematical description of the relationship between the transmitted signal and the observation at the receiver side; a channel \textit{simulator} is a software that exactly evaluates the channels in a specific scenario without seeking a mathematical description. Full-wave simulations or ray tracing can constitute very accurate channel simulators if enough care is taken in representing the specific propagation environment in the simulator. However, channel simulations, especially full-wave simulations, are often computationally expensive when targeting complex wireless scenarios, and do not offer mathematical insight into the channels. 

A wide variety of channel models (e.g., Rayleigh, Rice, Weibull, and Nakagami-$m$) are used to describe fading wireless environments through \textit{statistical} approaches~\cite{simon2005digital}. However, given the advent of ``smart radio environments,'' it becomes crucial to accurately include the \textit{deterministic} RIS-based parametrization of the wireless environment in the channel models. Clearly, such deterministic control is not compatible with the statistical philosophy of traditional channel models. Moreover, given the complexity of wave physics and its underlying principles, including long-range mesoscopic correlations~\cite{hsu2017correlation}, causality (a system's output cannot temporally precede its input), 
the notion of space (i.e., the spatial location of the involved entities), 
dispersion (i.e., frequency selectivity), the intertwinement of an RIS element's phase and amplitude response, and energy conservation, it is tremendously difficult to formulate accurate channel models for deterministic RIS-based control of wireless fading channels. 

Given the above-listed challenges, RIS-parametrized wireless channels have to date mainly been studied in free space without fading; we survey the three leading \textit{free-space} approaches in Sec.~\ref{subsec:Free}. Next, we survey attempts at marrying together the \textit{statistical} nature of random-matrix approaches to fading and the \textit{deterministic} nature of RIS-controlled wireless channels in Sec.~\ref{subsec:AdHoc}. We highlight in what aspects such \textit{ad hoc} modifications of random-matrix approaches are incompatible with wave physics. Then, we contextualize the coupled-dipole formalism on which PhysFad builds in Sec.~\ref{subsec:PhyFad}.

\subsection{Free Space RIS-Parametrized Channel Models}
\label{subsec:Free}

\subsubsection{Discrete Array of Mutually Independent Reflectors}
\label{subsub:convrismod}

The simplest and most common model discretizes the RIS into elements with a fixed reflection coefficient that does not depend on the configuration of neighboring RIS elements~\cite{ozdogan2019intelligent,bjornson2020power,najafi2020physics}. Dispersion (frequency-selectivity) and the influence of the angle of incidence are usually neglected. The reflection coefficients for different states of the RIS elements are often assumed to be arbitrarily tunable or to be $\pm1$, but their values can also be determined from equivalent circuit models~\cite{abeywickrama2020intelligent,li2021intelligent}, full-wave simulations, or experiments. For RISs with sub-wavelength elements, applying such a model typically requires grouping multiple RIS elements into macro-pixels. The applicability of such simple models has been confirmed with experimental prototypes~\cite{tang2020wireless,zhao2020metasurface}. Inter-element mutual coupling within this modeling framework is hence not always significant but can be accounted for as in Ref.~\cite{williams2020communication}.

\subsubsection{Inhomogeneous Surface-Impedance Sheet}

Assuming the RIS is a homogenizable metasurface (i.e., sub-wavelength unit cell dimensions and spacing), in free space the RIS can be modeled macroscopically in terms of a continuous surface impedance~\cite{danufane_PathLoss_2021,di2021communication}.
 
\subsubsection{Mutually Coupled Impedance-Modulated Antennas}
\label{subsub:fsmutcoup}

 By interpreting each RIS element as an impedance-modulated backscatter antenna~\cite{ivrlavc2010toward, alexandg_ESPARs}, the mutual coupling between transceiving antennas and all RIS elements can be rigorously formulated and related to the end-to-end channel matrix~\cite{gradoni_EndtoEnd_2020}.

\subsection{\textit{Ad Hoc} Modified Statistical Approaches}
\label{subsec:AdHoc}

The universality of random-matrix based models of fading channels originates from the fact that they require very little to no information about the wireless system that is being modeled. In other words, they are  agnostic to system-specific features; this property simplifies design and analysis, yet it becomes problematic once deterministic effects, such as a specific RIS configuration, are supposed to be taken into account. A representative example is the Rayleigh fading model, which has been extensively studied from a wave-physics perspective. Specifically, ``universal'' features of wave-chaotic fields inside complex scattering enclosures have been described as the superposition of a large number of plane waves with random phases, amplitudes, and directions~\cite{stockmann2000quantum,hemmady2005universal,yeh2012first,gradoni2014predicting}. Given the importance of non-universal deterministic effects, \textit{ad hoc} modifications of random-matrix approaches were explored in the frequency domain to include some features, such as direct paths~\cite{baranger1996short,kuhl2005direct,yeh2010experimental,savin2017fluctuations,del2020implementing}. Corresponding time-domain results are still under development~\cite{mapredicting}. 
Alternatively, geometry-based stochastic channel models, which statistically describe the explicit locations of scatterers, were proposed and adopted by popular channel simulators such as COST 2100~\cite{liu2012cost} and QuaDRiGa~\cite{jaeckel2014quadriga}.

It remains an open challenge how to incorporate the deterministic features of a specific RIS configuration in a random-matrix framework. The use of a series of \textit{random} RIS configurations is compatible with the ``universality'' of random-matrix models~\cite{del2020implementing}, however, the deterministic programming of the scattering environment with \textit{optimized} RIS configurations is not. Currently, the common strategy is to formulate the end-to-end channel matrix as $\mathbf{H} = \mathbf{H}_{RX-RIS}\mathbf{\Phi}_{RIS}\mathbf{H}_{TX-RIS}$, where $\mathbf{H}_{RX-RIS}$ and $\mathbf{H}_{TX-RIS}$ are 
random matrices emulating fading (e.g., Rayleigh) between the multi-antenna transceivers and the RIS, and $\mathbf{\Phi}_{RIS}$ contains the  reflection coefficients of the RIS elements~\cite{tahir2020analysis,ferreira2020bit,basar2021reconfigurable,selimis2021performance,Moustakas_RIS,zhang2021spatial,arslan2021over,sun2021channel,phan2022performance}. 
Such formulations represent an one-way cascade of multiple scattering events, followed by one interaction with the RIS and further multiple scattering events. 
However, in general, \textit{the RIS is an inseparable part of the scattering environment} and in the presence of strong multipath, rays typically bounce off the RIS multiple times, sabotaging any linear relationship between the RIS configuration and channel coefficients. 
Indeed, experiments showed that the impact of any given RIS element on the channel coefficients is in general not independent from the configuration of the other RIS elements~\cite{dupre2015wave}. This dependence does not originate from coupling between neighboring RIS elements, but from reverberation-induced long-range correlations: a given ray typically encounters multiple, not necessarily neighboring, RIS elements during its trajectory. This reverberation is a fundamental property of non-trivial scattering media that can be harnessed as a virtue: for instance, recent experiments demonstrated reverberation-assisted deeply sub-wavelength localization using an RIS and a single-antenna receiver~\cite{del2021deeply}.
Nonetheless, in the high-attenuation regime, experimental evidence suggests that a linear approximation based on a single interaction with the RIS and the absence of long-range mesoscopic correlations approximately holds~\cite{del2016intensity,Kaina_metasurfaces_2014}. However, the use of random matrices ($\mathbf{H}_{RX-RIS}$ and $\mathbf{H}_{TX-RIS}$) which lack any notion of space is incompatible with causality, which is a particularly pressing problem when working in the time domain (see Secs.~\ref{sec:TDRep}~and~\ref{sec:CaseStudy}).

\subsection{Channel Modeling via the Coupled-Dipole Formalism}
\label{subsec:PhyFad}
In this paper, we introduce PhysFad, a rigorous physics-based end-to-end channel model for RIS-parametrized wireless environments \textit{with adjustable fading}. The purpose of PhysFad is \textit{not} to \textit{simulate} a specific wireless setting, but to \textit{model} different types of RIS-parametrized fading channels while respecting all aspects of wave physics. 
Because PhysFad is derived from first principles, it naturally complies with the wave-physical reality in both frequency and time domains. 
PhysFad is built upon an exact analytical formulation. The underlying \acl{cdf} is conceptually related to frameworks of mutually coupled antennas~\cite{ivrlavc2010toward,alexandg_ESPARs,gradoni_EndtoEnd_2020}.
PhysFad's approach of decomposing the surfaces of the scattering environment into a discrete collection of dipoles is quite common in electromagnetism~\cite{draine1994discrete}. For instance, the equivalent surface current distribution of antennas is often modeled as collection of discrete dipoles~\cite{balanis2012advanced}, which is closely related to the method of moments~\cite{harrington1993MoM} and also resembles time-domain finite-element boundary integrals~\cite{jiao2001fast}. The \acl{cdf} has been used for decades to model light scattering~\cite{purcell1973scattering,mulholland1994light,de1998point}, and also for years in the metamaterials community~\cite{bowen2012using,f2016analytical,yves2017crystalline,Orazbayev2018,pulido2018analytical,del2020learned,yoo2019analytic}. Recently, the \acl{cdf} was employed to model slow and fast fading in rich scattering environments without RISs in Ref.~\cite{del2021deeply} and Ref.~\cite{LocalizationDynamicEnvironment}, respectively.

\section{The PhysFad Channel Model}
\label{sec:ChModel}
In this section, we introduce the PhysFad channel model, which employs a generic scalar 2D \acl{cdf} whose essential aspects are described in Sec.~\ref{subsec:ChModCDF}. 
PhysFad describes each of the three entities affecting the wireless channels, namely the transceiving antennas, the scattering environment, and the programmable RIS elements, as a dipole or a collection of dipoles with specific properties, as presented in Secs.~\ref{subsec:ChModTransceivers}-\ref{subsec:ChModRIS}. 
These components are combined into a channel model in terms of the end-to-end channel matrix in Sec.~\ref{subsec:ChModMatrix}, whose fading level can be adjusted as shown in Sec.~\ref{subsec:AdFad}. The overall channel model is summarized in Sec.~\ref{subsec:AlgSumm}. 

We develop PhysFad based on the 2D \acl{cdf}; future work can extend our formalism to a dyadic 3D version and determine the parameters such that they describe a specific type of a transceiver antenna, a specific RIS, and a specific wireless environment. Furthermore, specific devices can be modeled via the collective response of a collection of dipoles with suitable parameters~\cite{bertrand2020global} and/or by including multi-pole terms~\cite{lemaire1997coupled}. We leave such developments for future work; here, our goal is to describe a generic formalism that is representative of typical wireless communication scenarios. Without loss of generality, we will hence work with arbitrary units such that the central operating frequency as well as the medium's permittivity and permeability are all defined to be unity in the following.

\subsection{Coupled-Dipole Formalism}
\label{subsec:ChModCDF}
A dipole is a system which consists of a pair of charges of equal magnitude $q$, but opposite sign, that are separated by some distance $\delta$~\cite{novotny2012principles}. Working in 2D, we consider dipoles in the $x-y$ plane whose dipole moments are oriented along the vertical $z$ axis.
For concreteness, picture our 2D dipoles as vertical vias inside a parallel-plate waveguide of height $\delta$ which should be less than half a wavelength~\cite{pulido2018analytical}. Equivalently, by the image theorem, we can think of infinitely long vias in absence of the waveguide~\cite{pulido2018analytical}. The dipole moment $p(f)=q\delta= \frac{I}{\jmath(2\pi f)}\delta$ quantifies the dipole's polarity, where $\jmath\triangleq\sqrt{-1}$ and $I$ and $f$ denote current and frequency, respectively. 
The polarizability $\alpha$ quantifies the dipole's tendency to acquire a dipole moment when an electric field is applied. The dipole moment ${p_i}(f)$ of the $i$th dipole is  related to the local electric field at the dipole's position $\mathbf{r}_i$ via the dipole's frequency-dependent polarizability $\alpha_i(f)$:
\begin{table} 
\caption{Summary of key variables, indicating the utilized symbols and corresponding SI units.}
\vspace{-0.4cm}
\label{tbl:Symbols}
\begin{center}
\begin{tabular}{ |c|c|c|c| } 
\hline
Variable & Symbol & SI Units \\
\hline
Charge & $q$ & $\mathrm{C}=\mathrm{A}\cdot\mathrm{s}$ \\ 
Dipole Moment & $p$ & $\mathrm{C}\cdot\mathrm{m}=\mathrm{A}\cdot\mathrm{s}\cdot\mathrm{m}$ \\ 
Electric Field & $E$ & $\mathrm{V}\cdot\mathrm{m}^{-1} = \mathrm{A}^{-1}\cdot\mathrm{s}^{-3}\cdot\mathrm{kg}\cdot\mathrm{m}$ \\ 
Polarizability & $\alpha$ & $\mathrm{C}\cdot\mathrm{V}^{-1}\cdot\mathrm{m}^2 = \mathrm{A}^2\cdot\mathrm{s}^4\cdot\mathrm{kg}^{-1}$ \\ 
Charge Term & $\chi^2$ & $\mathrm{C}^2\cdot\mathrm{kg}^{-1} = \mathrm{A}^2\cdot\mathrm{s}^2\cdot\mathrm{kg}^{-1}$ \\
Resonance Frequency & $f_{\mathrm{res}}$ & $\mathrm{s}^{-1}$\\
Radiation Damping & $\gamma^R$ & $\mathrm{s}^{-2}$\\
Absorptive Damping Term & $\Gamma^L$ & $\mathrm{s}^{-1}$\\
Free Space Green's Function & $G$ & $\mathrm{C}^{-1}\cdot\mathrm{V}\cdot\mathrm{m}^{-2} = \mathrm{A}^{-2}\cdot\mathrm{s}^{-4}\cdot\mathrm{kg}$ \\
Wavenumber & $k$ & $\mathrm{m}^{-1}$ \\
Permittivity & $\epsilon$ & $\mathrm{F}\cdot\mathrm{m}^{-1} =  \mathrm{A}^2\cdot\mathrm{s}^4\cdot\mathrm{kg}^{-1}\cdot\mathrm{m}^{-3}$ \\
Dipole Size & $\delta$ & $\mathrm{m}$ \\
\hline
\end{tabular}
\end{center}
\end{table}
\begin{equation}
{p_i}(f) = \alpha_i(f) {E_{\text{loc}}}(\mathbf{r}_i,f).
\label{eq:p}
\end{equation}
We use the following Lorentzian model for the polarizability~\cite{novotny2012principles}: 
\begin{equation}
\alpha_i(f) = \frac{\chi_i^2}{4\pi^2f_{\text{res},i}^2 - 4\pi^2f^2 + \jmath (\gamma_i^R + 2\pi f\Gamma_i^L) },
\label{eq:alpha}
\end{equation}
where the charge term $\chi_i^2$ has dimensions of the square of a charge over the charge's mass and acts like an amplitude term of $\alpha_i(f)$, $f_{\mathrm{res}}$ is the resonance frequency, $\gamma_i^R$ denotes inevitable radiation damping, and $\Gamma_i\geq 0$ is the absorptive damping term. Energy conservation requires $\textrm{Im}(\alpha^{-1}(f))\geq\frac{\gamma_i^R}{\chi_i^2}=\frac{k^2}{4\epsilon \delta}$~\cite{sipe1974macroscopic,strickland2015dynamic}, where $\epsilon$ denotes the permittivity and $k$ is the wavenumber.

The local field ${E_{\text{loc}}}$ at the $i$th dipole is the superposition of the external field ${E_{\text{ext}}}$ exciting the system and the fields radiated by the other dipoles: 
\begin{equation}
{E_{\text{loc}}}(\mathbf{r}_i,f) = {E_{\text{ext}}}(\mathbf{r}_i,f) + \sum_{j \neq i}G_{ij}\left( \mathbf{r}_i,\mathbf{r}_j,f\right){p_j}(f).
\label{eq:Eloc}
\end{equation}
In Eq.~\eqref{eq:Eloc}, the contribution of $p_j$ to ${E_{\text{loc}}}(\mathbf{r}_i,f)$ is weighted by
\begin{equation}
G_{ij}\left( \mathbf{r}_i,\mathbf{r}_j,f\right) = -j \frac{k^2}{4\epsilon \delta} \text{H}_0^{(2)}\left( k \left| \mathbf{r}_i - \mathbf{r}_j \right|\right)
\label{eq:G}
\end{equation}
which represents the 2D free-space Green's function between the positions $\mathbf{r}_i$ and $\mathbf{r}_j$ with $\text{H}_0^{(2)}(\cdot)$ denoting a Hankel function of the second kind~\cite{harrington1961time}.\footnote{The electric field induced at $\mathbf{r}_i$ by a unit dipole moment $p_j(f)$ at $\mathbf{r}_j$ in free space is $G_{ij}\left( \mathbf{r}_i,\mathbf{r}_j,f\right)p_j(f)$.} Substituting Eq.~\eqref{eq:p} into Eq.~\eqref{eq:Eloc}, we obtain
\begin{equation}
\alpha_i^{-1}(f){p_i}(f) - \sum_{j \neq i}G_{ij}\left( \mathbf{r}_i,\mathbf{r}_j,f\right){p_j}(f) = {E_{\text{ext}}}(\mathbf{r}_i,f),
\label{eq:equation}
\end{equation}
which can be solved for the dipole moments via matrix inversion at each considered frequency (see Sec.~\ref{subsec:ChModMatrix}). The symbols used above and their SI units are summarized in Table~\ref{tbl:Symbols}.

Having recalled the well-established 2D \acl{cdf}, we now relate the essential parameters of each dipole ($\mathbf{r}_i,\ f_{\text{res},i},\ \chi_i,\ \Gamma_i^L$) to its role in the wireless communication system.

\subsection{Modeling of Transceivers}
\label{subsec:ChModTransceivers}
\begin{figure}
    \centering
    \includegraphics[width=0.6\columnwidth]{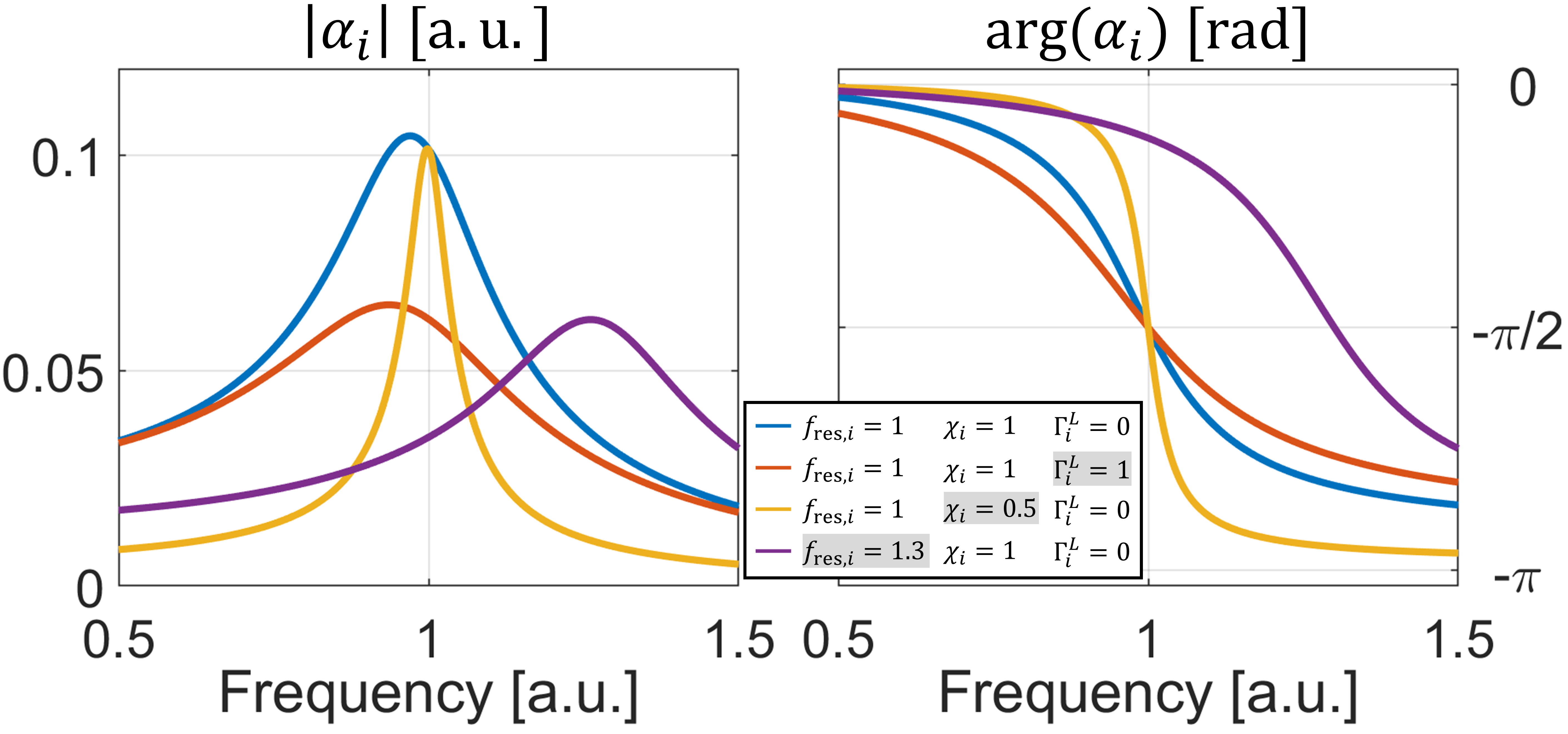}
    \caption{Impact of parameters $\Gamma^L_i$, $\chi_i$, and $f_{\text{res}_i}$ on magnitude and phase of the Lorentzian polarizability $\alpha_i(f)$ defined in Eq.~\eqref{eq:alpha}.}
    \label{fig:Lorentzian}
\end{figure}

The first ingredient are the transceivers that generate and capture the waves that carry the  information. In most wireless communication systems, the utilized antennas are resonant within the operated frequency band. A convenient example is a half-wave dipole made up of thin wires which is naturally well-described by a dipole resonant at the central operating frequency. Crucial antenna properties, such as central operating frequency and bandwidth, can be adjusted via $f_{\text{res},i}$, $\chi_i$, and $\Gamma_i^L$ as shown in Fig.~\ref{fig:Lorentzian}. Non-resonant antenna properties that are essentially flat within the considered frequency band can be obtained by choosing $f_{\text{res},i}\gg 1$.
In order to act as transmitter, we impose a desired non-zero external electric field at the transmitting dipole's location (see also Sec.~\ref{subsec:ChModMatrix}). The external field is zero everywhere except at the transmitting dipoles' locations. 

The above discussed options model a ``field-invasive'' transceiver that inevitably scatters waves. Although less common, the 2D \acl{cdf} can also accommodate non-scattering transceivers. A non-scattering transmitter can be described without associated dipole simply through an external field. For instance, a non-invasive point source at location $\mathbf{r}_t$ results in an external field ${E_{\text{ext}}}(\mathbf{r}_i,f)= \frac{E_0}{\alpha_0} G_ {it}(\mathbf{r}_i,\mathbf{r}_t,f)$, where $\frac{E_0}{\alpha_0}$ describes how strongly the source emits. A non-invasive receiver at location $\mathbf{r}_r$ is similarly modeled without dipole simply by evaluating the local electric field at $\mathbf{r}_r$ using Eq.~\eqref{eq:Eloc}.

\subsection{Modeling of the Scattering Environment}
\label{subsec:ChModEnvironment}

Having covered the transceivers, we now turn our attention to the wireless environment (excluding the RIS which is covered in the next subsection). 
For free-space scenarios, the scattering environment is trivial, as in most papers on RIS-parametrized channel modeling (Sec.~\ref{subsec:Free}). But wireless communication is often concerned with non-trivial dynamic scattering environments that give rise to fading. To fully explore and reap the potential of RIS-parameterization in wireless communication, the ability to model fading in a physically justified and adjustable manner is thus crucial. This subsection gives a brief overview on how to use the \acl{cdf} to introduce a scattering environment. We provide a specific example of implementation and characterization of adjustable Rician fading in Sec.~\ref{subsec:AdFad}.

\begin{figure*}
    \centering
    \includegraphics[width=15cm]{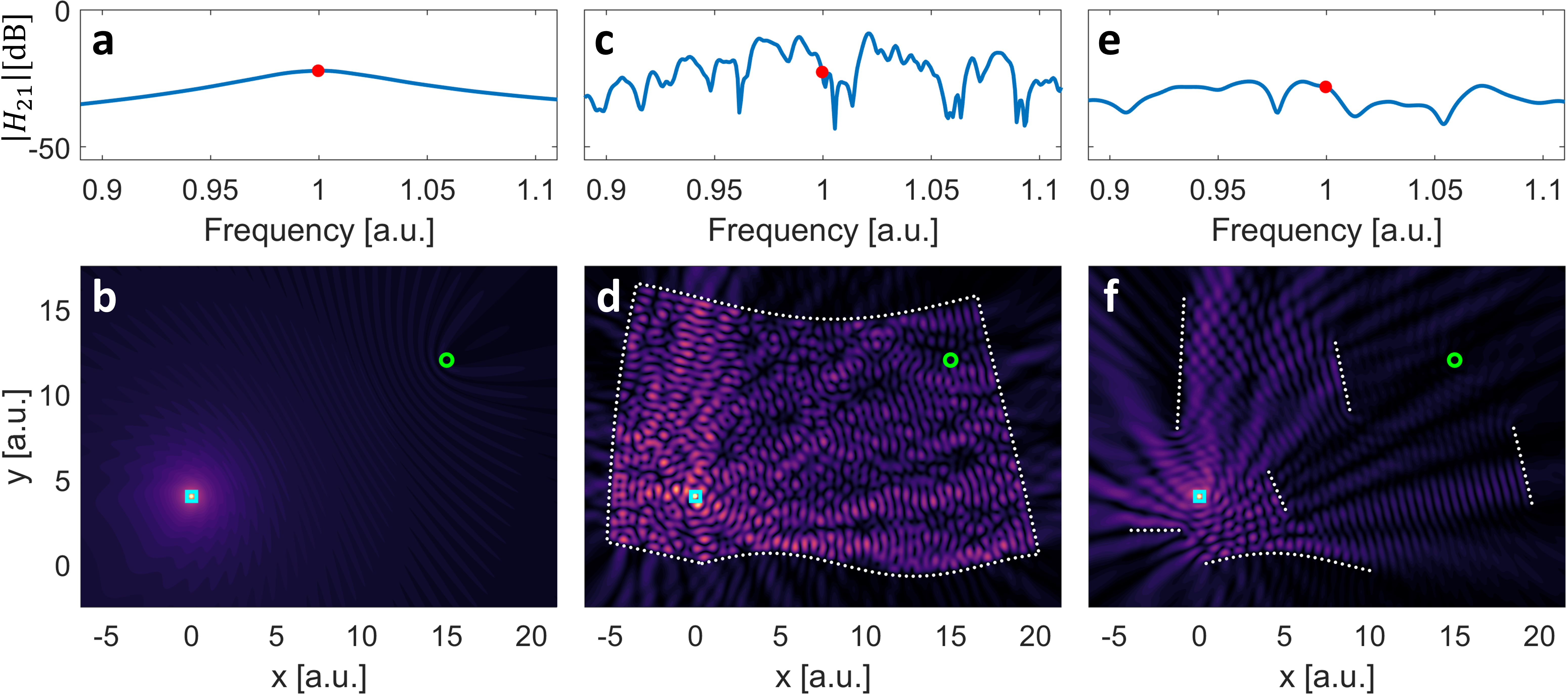}
    \caption{Illustration of the impact of various scattering environments on the transmission spectrum's magnitude $|H_{21}|$ (top row) and spatial field distribution (bottom row, field magnitude plotted at the central frequency indicated by a red dot in the top row). The utilized parameters are \{$\Gamma^L=0$, $\chi=0.5$, $f_{\text{res}}=1$\} for the transceiver dipoles, and \{$\Gamma^L=0$, $\chi=50$, $f_{\text{res}}=10$ \} for the fence dipoles. The spatial fields are evaluated using Eq.~\eqref{eq:Eloc}. The dipole locations are indicated through white dots on top of the spatial field maps. The three considered scattering environments are: free space (left); metallic electrically-large irregularly shaped enclosure (middle); disordered collection of metallic obstacles in free space, e.g., an outdoor environment (right). }
    \label{fig:BasicFieldPlots}
\end{figure*}

The first challenge is hence to introduce a scattering environment. Wave propagation in a static scattering environment yields multipath links that is at the origin of \textit{slow} fading. 
As noted above, we decompose scattering surfaces into discrete collections of dipoles. For instance, in Refs.~\cite{LocalizationDynamicEnvironment,del2021deeply} this technique was used to represent electrically large metallic enclosures which are of direct relevance to wireless communication inside vessels, planes, trains, or busses. These metallic enclosures were modeled as a dense fence of dipoles whose resonance frequency lies well above the considered frequency band. The latter guarantees that the properties of the enclosure are roughly the same at all considered frequencies. Fence density and losses in the fence dipoles allow one to adjust the amount of reflection and absorption by the fence, respectively. Similarly, other complex propagation environments such as a outdoor settings with a multitude of reflecting objects surrounding the transceivers can be implemented. Illustrative examples of different classes of scattering environments are shown in Fig.~\ref{fig:BasicFieldPlots}, revealing fundamental differences between trivial wave propagation in free space and rich scattering inside enclosures or collections of obstacles. The spatial field distribution in Fig.~\ref{fig:BasicFieldPlots}d is known as speckle pattern and arises from the interference of countless waves reflected off the enclosure's walls. One can also include resonant scatterers in the wireless environment by utilizing dipoles whose resonance frequencies are chosen to lie within the operating band. The latter is the basis of our RIS model (Sec.~\ref{subsec:ChModRIS}).

The second challenge is to add dynamic effects such that the scattering environment changes rapidly (corresponding to a short channel coherence time), giving rise to \textit{fast} fading. This can be conveniently achieved by evaluating the channel (see Sec.~\ref{subsec:ChModMatrix}) for multiple variations of one or multiple scattering objects (see Sec.~\ref{subsec:AdFad}). A simple example in Ref.~\cite{LocalizationDynamicEnvironment} involved one metallic object that rotates around its own axis and is at an arbitrary angular position at any given instant in time, emulating, e.g., a rotating fan.

\subsection{RIS Element Modeling and Characterization}
\label{subsec:ChModRIS}

Having clarified how transceivers and the scattering environment (excluding RISs) can be incorporated into the \acl{cdf}, we now focus on RISs. An RIS is an array of elements with programmable scattering properties (usually in reflection). The vast majority of RIS prototypes relies on meta-atoms with programmable resonances. For instance, the RIS element from Ref.~\cite{Kaina_metasurfaces_2014} uses a PIN diode whose bias voltage controls whether the meta-atom is resonant or not at the operating frequency. A physically faithful RIS model should account for: \textit{i}) the intertwinment of amplitude and phase in a typical Lorentzian resonator; \textit{ii}) the frequency-dependence of a typical Lorentzian resonator; and \textit{iii}) the coupling effects between nearby RIS elements. 
Multiple recent papers on ``electromagnetics-compliant'' RIS models have discussed how to account for some or all of these aspects in free space~\cite{williams2020communication,li2021intelligent,abeywickrama2020intelligent,gradoni_EndtoEnd_2020}. 

A physically faithful basic RIS element model consists hence of one dipole whose resonance frequency can be changed depending on the desired configuration. For a $1$-bit programmable RIS, we simply switch between a resonance frequency at the center of the considered frequency band and a resonance frequency well outside the considered frequency band in order to emulate the two possible states. Multi-bit or continuous tuning of RIS elements can, of course, be implemented through a more fine-grained control of the RIS element's resonance frequency. Our description of the RIS imposes no limitations on the spatial arrangement of the RIS elements, and can readily be applied to conformal or distributed RIS prototypes. The scattering properties of RIS elements can also be programmed mechanically as opposed to electrically; for instance, Ref.~\cite{mechanical_RIS} experimentally presented an array of metal blocks with adjustable height. An interesting feature of this unconventional design is its broadband non-Lorentzian nature because the design is not based on a resonant phenomenon. PhysFad can accommodate such RIS designs by describing each RIS element as one (or multiple) non-resonant dipole(s) whose location(s) is (are) physically adjusted according to the desired RIS configuration.

\begin{figure}
    \centering
    \includegraphics[width=\columnwidth]{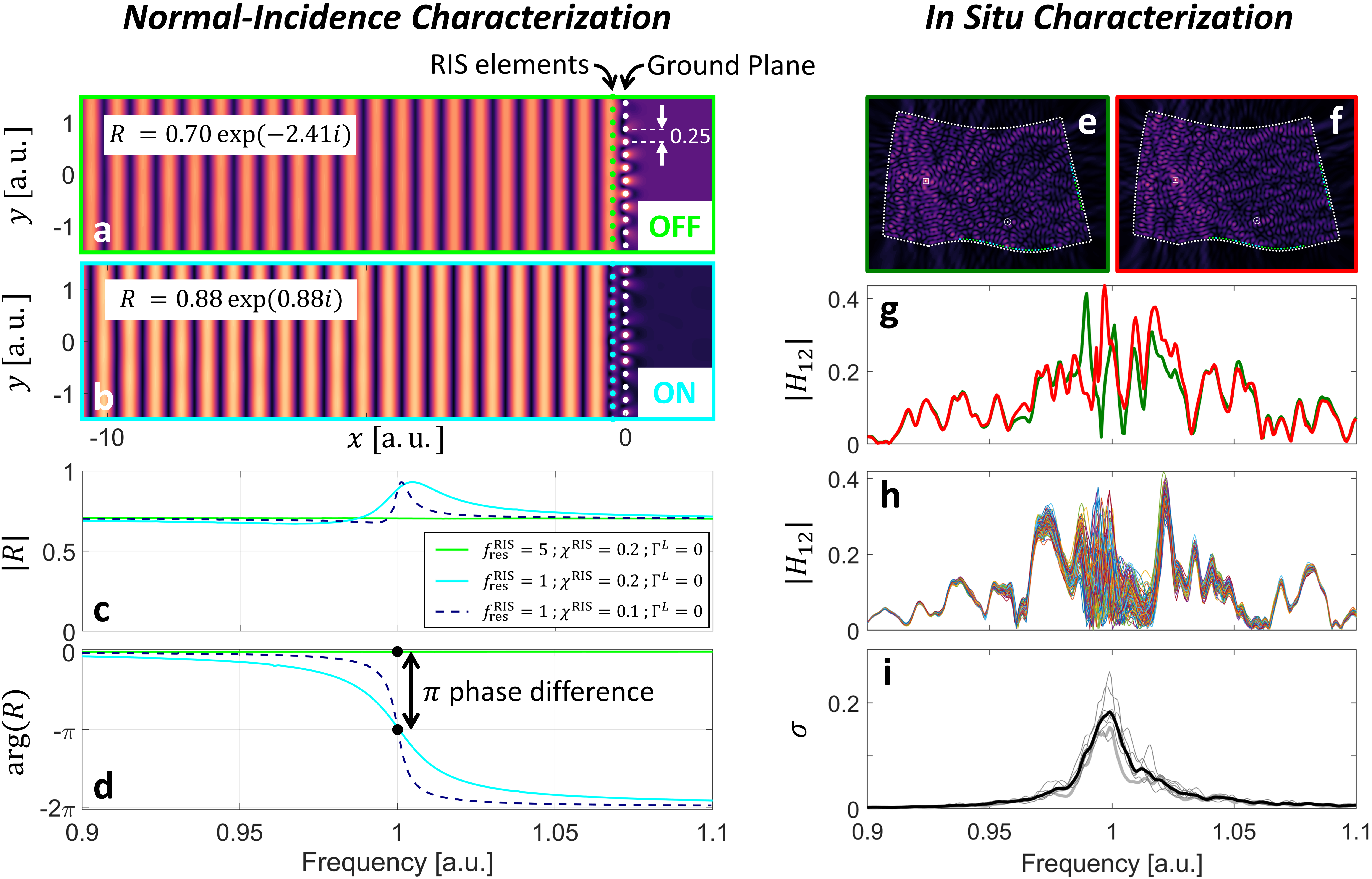}
    \caption{Characterization of a $1$-bit programmable RIS under normal plane wave incidence (left) and \textit{in situ} (right). 
    For the former, the setup involves a ground plane (white dipole fence with \{$\Gamma^L=0$, $\chi=50$, $f_{\text{res}}=10$\} and dipole separation $d_w = 0.25$) and a series of programmable RIS elements (green/cyan-colored dipoles with \{$\Gamma^L=0$, $\chi=0.2$, $f_{\text{res}}\in\{1,5\}$\}, separation $d_w = 0.25$, and $\Delta = 0.25$ in front of the ground plane). The 1-bit programmable RIS elements are either OFF resonance ($f_{\text{res}}^{\rm RIS}=5$, green) or ON resonant ($f_{\text{res}}^{\rm RIS}=1$, cyan). For $f=1$, field magnitude plots for the two cases in a and b reveal standing wave patterns which result from the superposition of a plane wave travelling to the right and a reflected plane wave travelling to the left. We extract from these standing wave pattern the reflection coefficient $R$ under normal incidence at each frequency. 
    Magnitude and phase of $R$ are plotted as a function of frequency in c and d, respectively. The phase difference at $f=1$ reaches exactly the desired value of $\pi$. Moreover, in c and d we also display the reflection coefficient of the ON state for RIS elements with $\chi^{\rm RIS}=0.1$ (dashed dark blue).
    For the \textit{in situ} characterization, we consider a distributed conformal $45$-element $1$-bit-programmable RIS inside the complex scattering enclosure from Fig.~\ref{fig:BasicFieldPlots}d. Field magnitude maps for two random RIS configurations (ON [OFF] elements shown in green [cyan]) at $f=1$ are displayed in e and f. The magnitude of the transmission spectrum $H_{12}$ between the two transeivers for the two cases is shown in g. The same quantity for $100$ random RIS configurations is shown in h. The standard deviation $\sigma$ of the complex-valued transmission spectra $H_{12}$ across random RIS configurations is shown as gray thick line in i. Other gray thin lines show $\sigma$ for different randomly chosen transceiver locations. The average over these realizations of transceiver locations is shown as black thick line in i.}
    \label{fig:RisCharac}
\end{figure}

We now characterize our basic RIS design in the conventional manner by evaluating the reflection coefficient $R$ for various RIS configurations under normal incidence. In Fig.~\ref{fig:RisCharac}(a,b) we emulate a plane wave normally incident on an ``infinitely'' large $1$-bit programmable RIS in which all meta-atoms are in the same state (ON or OFF); then, we extract $R$ for normal incidence by fitting the ensuing standing wave pattern. At the central operating frequency, our RIS design displays a phase difference of exactly $\pi$ (Fig.~\ref{fig:RisCharac}(d)). Our results also show the expected frequency selectivity of the RIS that can be tuned, for instance, through the parameter $\chi^{\rm RIS}$. 
We see in Fig.~\ref{fig:RisCharac}(c) that off resonance $|R|=0.70$; this value increases to $0.88$ if a five times denser dipole fence serves as ground plane. At resonance (at $f=1$ for $f_{\mathrm{res}}^{\mathrm{RIS}} = 1$), $|R|$ is slightly higher because there are essentially two barriers that prevent transmission toward the right side and we have set the absorptive damping term $\Gamma^L$ to zero for the RIS elements.

While we consider a planar RIS in Figs.~\ref{fig:RisCharac}(a-d), PhysFad is capable of simulating conformal and/or distributed RISs, as seen in Figs.~\ref{fig:RisCharac}(e,f). In Figs.~\ref{fig:RisCharac}(e-i) we characterize \textit{in situ} a $45$-element $1$-bit-programmable RIS whose properties are similar to the RIS from Figs.~\ref{fig:RisCharac}(a,b), but which is distributed in a conformal manner across two parts of the walls of the complex scattering enclosure from Fig.~\ref{fig:BasicFieldPlots}d. Specifically, we characterize the ability of the RIS to modulate the field, accounting for all possible angles of incidence~\cite{Kaina_metasurfaces_2014,alexandg_2021}. The field magnitude maps for two random RIS configurations at $f=1$ are seen in Fig.~\ref{fig:RisCharac}(e,f) to differ, and this observation becomes clearer upon inspecting the magnitude of the transmission spectrum between the two transceivers as a function of frequency in Fig.~\ref{fig:RisCharac}g. The two curves differ strongly in the vicinity of $f=1$ where the RIS efficiently modulates the field, but are almost identical at frequencies further away from $f=1$. This observation relates once again to the frequency-selective nature of resonance-based RIS designs. The bandwidth of the considered transceivers is also clearly seen once again as a global envelope over the chaotic transmission spectrum. To further visualize at which frequencies the transmission varies the most if random RIS configurations are applied, we superpose $100$ such curves in Fig.~\ref{fig:RisCharac}h. 

The \textit{in situ} characterization~\cite{Kaina_metasurfaces_2014,alexandg_2021} consists in evaluating at each frequency the standard deviation $\sigma$ of the complex-valued transmission coefficient across a series of random RIS configurations. If the RIS efficiently modulates the field at a given frequency, $\sigma$ will be high. The resulting curve is plotted as thick gray line in Fig.~\ref{fig:RisCharac}i. However, this characterization is still specific to the choice of transceiver locations. To clearly extract the characteristics of the RIS, we average out this dependence by repeating the above procedure for multiple randomly chosen transceiver locations within the enclosure, and averaging over the corresponding standard deviations. The resulting curve (black line in Fig.~\ref{fig:RisCharac}i) is the result of our \textit{in situ} characterization and the identified operational bandwidth (frequency selectivity) confirms the one previously obtained result in Fig.~\ref{fig:RisCharac}(c,d) for normal incidence.

\subsection{End-to-End Channel Modeling}
\label{subsec:ChModMatrix}
We are now in a position to bring together all the ingredients of PhysFad in order to identify the end-to-end channel model. We formulate PhysFad in terms of the electric field $E_{\mathrm{loc},i}(\mathbf{r}_i,f)$ at the $i$th transceiver location $\mathbf{r}_i$, which is directly proportional to the current $I_i$ and voltage $V_i$ across this transceiver:
%
\begin{subequations}
    \begin{align}
I_i(f)&=\frac{j\omega p_i(f)}{\delta}= \frac{\jmath\omega \alpha_i(f)}{\delta} E_{\mathrm{loc},i}(\mathbf{r}_i,f),\\
V_i(f)&=Z_i I_i(f) = \frac{\jmath\omega \alpha_i(f) Z_i}{\delta} E_{\mathrm{loc},i}(\mathbf{r}_i,f),
    \end{align}
\label{eq:Z}
\end{subequations}

\noindent where $Z_i$ denotes the load impedance at the transceiver. 
We refer in the following to the input-output relation between electric fields at $N_{\rm R}$ receiving antennas and $N_{\rm T}$ transmitting antennas as the $N_{\rm R}\times N_{\rm T}$ complex-valued channel matrix $\mathbf{H}$, because in our case of identical transceivers this is directly proportional to the usual definition as the input-output relation in terms of voltages. 

Let us consider the most general MIMO scenario involving $N_{\rm T}$ transmitters and $N_{\rm R}$ receivers (Sec.~\ref{subsec:ChModTransceivers}), $N_{\rm E}$ dipoles that constitute the scattering environment (Sec.~\ref{subsec:ChModEnvironment}), and $N_{\rm RIS}$ dipoles that constitute the RIS (Sec.~\ref{subsec:ChModRIS}). Following the \acl{cdf} (Sec.~\ref{subsec:ChModCDF}), we note that the total number of dipoles in our system is $N\triangleq N_{\rm T}+N_{\rm R}+N_{\rm E}+N_{\rm RIS}$, and we begin by rewriting Eq.~\eqref{eq:equation} in matrix form:
\begin{equation}
\mathbf{W}(f) \mathbf{p}(f) = \mathbf{E}_{\rm ext}(f),
\label{eq:matrix_equation}
\end{equation}
where the vector $\mathbf{p}(f) = [p_1(f)\,p_2(f)\,\cdots\,p_N(f)]$ contains the dipole moments of our $N$ dipoles at frequency $f$, and $\mathbf{E}_{\rm ext}(f) = [E_{\text{ext},1}(f)\,E_{\text{ext},2}(f)\,\cdots\,E_{\text{ext},N}(f)]$ is comprised of the corresponding external electric fields. 
For the sake of readability, we are now printing $E_{\text{ext}}(\mathbf{r}_n,f)$ as $E_{\text{ext},n}(f)$. The $N\times N$ complex-valued matrix $\mathbf{W}(f)$ contains the inverse polarizabilities $\alpha_i^{-1}(f)$ defined in Eq.~\eqref{eq:alpha} of our $N$ dipoles along its diagonal (see  for the analytical expression), and the $(i,j)$th off-diagonal entry is $-G_{ij}(f)$, i.e., the negative of the 2D free-space Green's function between the locations of the $i$th and $j$th dipoles (see Eq.~\eqref{eq:G} for the analytical expression).

\begin{figure}[h]
    \centering
    \includegraphics[width=\columnwidth]{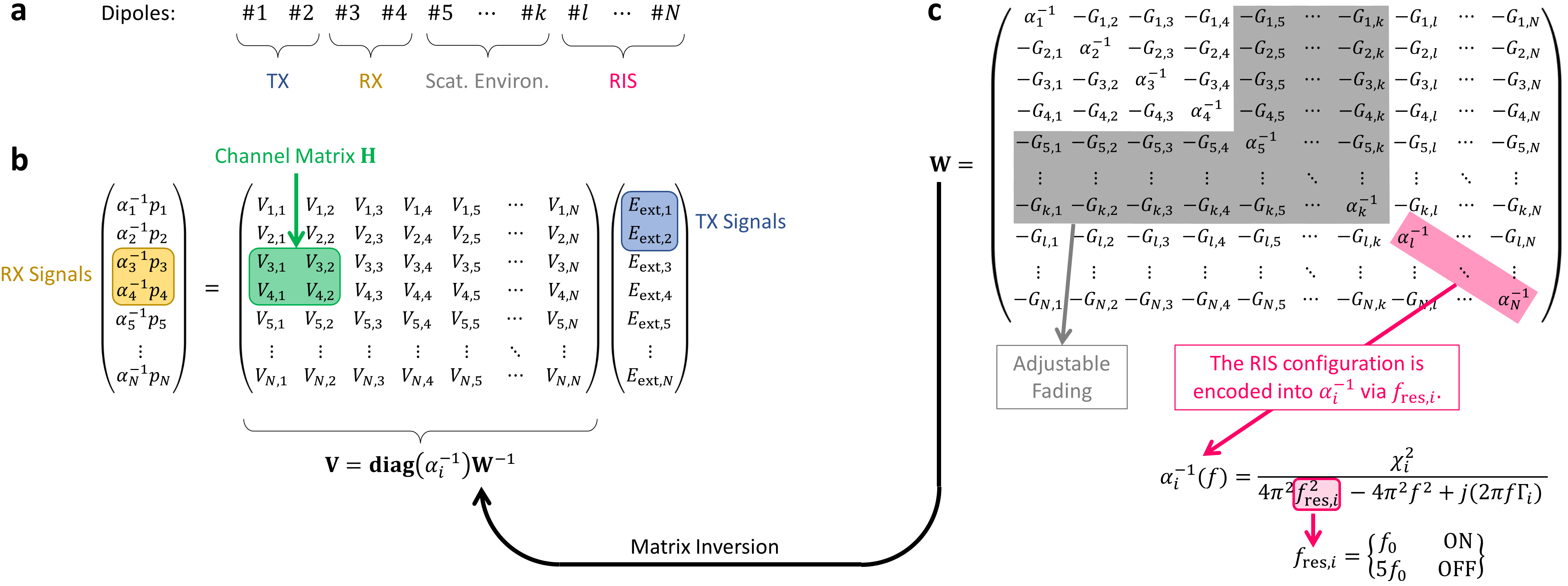}
    \caption{Illustration of the link between the \acl{cdf} and the RIS-parametrized end-to-end channel matrix for a $2\times2$ MIMO example. 
    a) Order of dipole indexing (without loss of generality). 
    b) Illustration of Eq.~\eqref{eq:final_inverted_matrix_equation} for the considered example, highlighting the part of $\mathbf{V}$ that is the sought-after end-to-end channel matrix $\mathbf{H}$, see also Eq.~\eqref{eq:channel_matrix}. The external field is zero for all but the transmitting dipoles, i.e., $E_{\mathrm{ext},i}=0$ $\forall$ $i>N_{\rm T}$.
    c) Parameterization of $\mathbf{W}$ through a binary RIS configuration. The RIS configuration dictates the resonance frequencies of the dipoles (indices from $l$ to $N$, see a) representing the RIS, and thereby their inverse polarizability. Ultimately, these inverse polarizabilities appear along the diagonal of $\mathbf{W}$. In addition, the adjustable fading is implemented via the part of $\mathbf{W}$ shaded in gray: the position of the scatterers via the off-diagonal gray entries, and their ``transparency'' (see Sec.~\ref{subsec:AdFad}) via the diagonal gray entries.}
    \label{fig:MatrixFigure}
\end{figure}

The wave equation's linearity allows us to perform our calculations independently at each desired frequency, implying great potential for parallelizing the evaluation of $\mathbf{H}$ and allowing us to henceforth drop the frequency dependence.
Recall that we know $\mathbf{W}$ and $\mathbf{E}_{\rm ext}$, and our goal is to compute $\mathbf{p}$ whose entries for the receiving dipoles must be multiplied by the corresponding inverse polarizabilities to obtain the received fields (see Eq.~\ref{eq:alpha}). Using standard matrix-inversion techniques, we thus first invert $\mathbf{W}$ to evaluate $\mathbf{p}$, yielding
\begin{equation}
\mathbf{p} = \mathbf{W}^{-1} \mathbf{E}_{\rm ext}.
\label{eq:inverted_matrix_equation}
\end{equation}
Next, we multiply both sides of Eq.~\eqref{eq:inverted_matrix_equation} by ${\rm diag}([\alpha_i^{-1}\,\alpha_2^{-1}\,\ldots\,\alpha_N^{-1}])$, which is a diagonal matrix containing the inverse polarizabilities:
\begin{equation}
{\rm diag}([\alpha_i^{-1}\,\alpha_2^{-1}\,\ldots\,\alpha_N^{-1}])\mathbf{p} = \mathbf{V} \mathbf{E}_{\rm ext},
\label{eq:final_inverted_matrix_equation}
\end{equation}
where we introduce $\mathbf{V}\triangleq{\rm diag}([\alpha_i^{-1}\,\alpha_2^{-1}\,\ldots\,\alpha_N^{-1}])\mathbf{W}^{-1}$. 
The end-to-end channel matrix $\mathbf{H}$ is now simply the $N_{\rm R}\times N_{\rm T}$ portion of the $N\times N$ matrix $\mathbf{V}$ that links the $N_{\rm T}$ transmitting dipoles to the $N_{\rm R}$ receiving dipoles. Without loss of generality, let us assume that the dipole indices are in the following order: transmitters, receivers, scattering environment, and RIS. Then,
\begin{equation}
\mathbf{H} = \mathbf{V}[(N_{\rm T}+1):(N_{\rm T}+N_{\rm R}),1:N_{\rm T}].
\label{eq:channel_matrix}
\end{equation}
An illustration for a $2\times 2$ MIMO example is provided in Fig.~\ref{fig:MatrixFigure}. Recall that the external field is zero for all but the transmitting dipoles, i.e., $E_{\mathrm{ext},i}=0$ $\forall$ $i>N_{\rm T}$.

\begin{rem}
Being derived from first principles, PhysFad inherently complies with \textit{all} aspects of wave physics and hence does \textit{not} require any \textit{ad hoc} corrections that are commonly used in unphysical channel models to ensure compliance with specific physical properties such as pathloss. 
If all dipoles constituting the scattering environment are removed, PhysFad collapses to the free-space channel model of RIS as mutually coupled impedance-modulated antennas from Sec.~\ref{subsub:fsmutcoup}~\cite{gradoni_EndtoEnd_2020}. Moreover, the conventional free-space channel model of an RIS as a discrete array of mutually independent reflectors (Sec.~\ref{subsub:convrismod}) can be obtained from PhysFad after multiple simplifying assumptions. Considering a SISO case \textit{in free space}, the received electric field is $E_r = \sum_{i \in  \mathrm{RIS}}G_{ir}p_i + G_{tr}p_t$, where the last term is the LOS contribution. Assuming that the local field at the transmitter is dominated by the imposed external field and assuming negligible coupling between RIS elements, $p_t\approx\alpha_t E_{\mathrm{ext},t}$ and $p_i\approx\alpha_i G_{it} p_t$, yielding $E_r\approx \left(\sum_{i \in  \mathrm{RIS}}G_{ir}\alpha_i G_{it} + G_{tr}\right)\alpha_t E_{\mathrm{ext},t}$, where $\alpha_i$ encodes the configuration of the $i$th RIS element (see Fig.~\ref{fig:MatrixFigure}(c)). 
\end{rem}

\subsection{Adjustable Fading}
\label{subsec:AdFad}
In the previous subsections, we established PhysFad's end-to-end channel model. Now, we explore in depth PhysFad's ability to implement adjustable fading. Because fading is a property of the uncontrolled part of the scattering environment (Sec.~\ref{subsec:ChModEnvironment}), we leave the RIS aside in this section. We illustrate how the tuning of a single parameter in a PhysFad model faithfully yields any desired Rician fading statistics. 

\begin{figure}[!t]
    \centering
    \includegraphics[width=0.9\columnwidth]{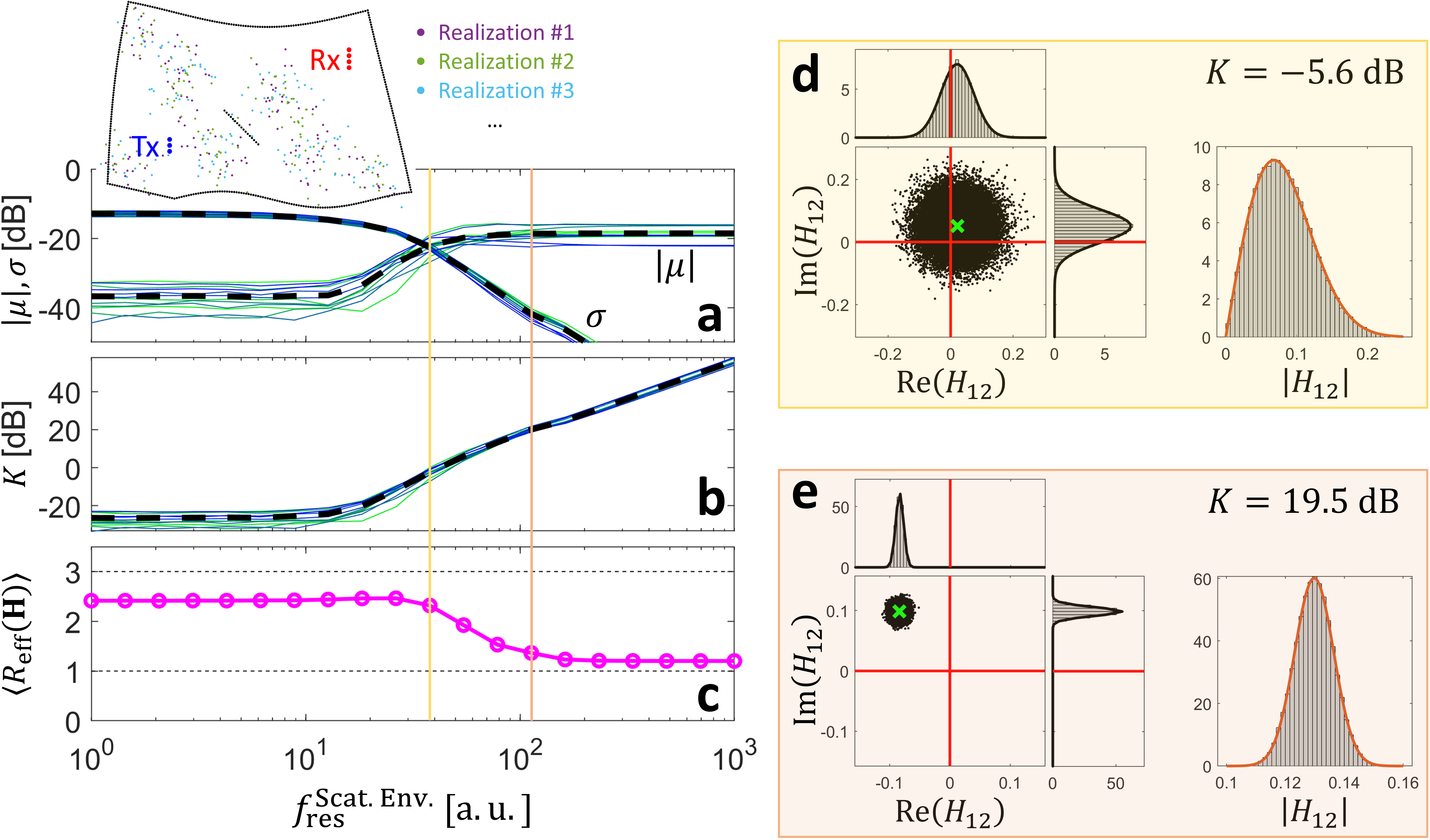}
    \caption{Illustration of adjustable Rician fading in a simple PhysFad model. The presented statistics are based on $5\times 10^4$ realizations; the dipole setup for the first three realizations is shown with color code as inset in the top left corner.
    We show the dependence of $|\mu|$ and $\sigma$ (a), $K$ (b), and $\langle R_{\mathrm{eff}}(\mathbf{H})\rangle$ (c) on the resonance frequency of the dipoles constituting the scattering environment. For a selected channel and two $K$-factor values, we show in (d,e) the cloud of channel coefficients in the complex plane (the green cross indicates the average). Probability density functions of the real and imaginary parts and the magnitude are shown with bars and fitted with Gaussian (real, imaginary) or Rician (magnitude) functions.}
    \label{fig:RiceFigure}
\end{figure}

It is well established that complex scattering enclosures (see Fig.~\ref{fig:BasicFieldPlots}d), also known as reverberation chambers (RCs), are ideally suited to emulate a radio environment with Rician fading~\cite{holloway2006use,corona2000reverberating,lemoine2010k,yeh2012first}. 
In Rician environments, the $K$-factor determines the relative strength of direct and scattered paths between a transmitter and a receiver. The Rayleigh environment is a special case of the Rician environment in which the direct contribution is negligible. In RC-based emulations of Rician environments, a wireless device is exposed to a statistical ensemble of fields by rotating a large irregularly shaped metallic object -- the so-called ``mode-stirrer'' -- inside the RC~\cite{holloway2006use,yeh2012first}. This mode-stirrer is easily implemented as part of the scattering environment in PhysFad~\cite{LocalizationDynamicEnvironment}. The $K$-factor is defined as \cite{simon2005digital}
\begin{equation}
K(f) = \frac{|\mu(f)|^2}{2\left[\sigma(f)\right]^2},
\label{eq:K}
\end{equation}
where $\mu(f)$ is the average and $\sigma(f)$ is the standard deviation of the ensemble of complex-valued transmission $H_{12}$ between two antennas. This definition is best understood by plotting the measured transmission values in the complex plane (see Fig.\ref{fig:RiceFigure}(d,e) for examples). Thereby, we find an approximately circular cloud of points that is centered off the origin; $|\mu|$ is the distance between the cloud's center and the origin, and $\sigma$ is the cloud's radius. Hence, Eq.~\eqref{eq:K} compares the direct and scattered intensities. The ``direct''  component is often referred to as LOS, although, to be precise, we note that it represents the interference of all paths that are static, i.e., paths that are not affected by the mode-stirrer rotations~\cite{yeh2012first}. In many practical scenarios, the LOS path dominates the static contribution but, even with blocked LOS, the cloud is typically centered off the origin (i.e., there is a non-LOS (NLOS) static contribution).

In RC-based experiments emulating Rician environments, it is possible to finely adjust the value of $K$ by using directive antennas and tuning their orientation, or by tuning the amount of absorption of the RC~\cite{holloway2006use,yeh2012first}. The former limits the applicability to directive antennas, excluding the use of omnidirectional antennas such as dipoles or small antennas. 
In our simulations, we can make use of a convenient tuning knob not available to the experimentalist: we can adjust the transparency of the environment via the resonance frequency $f_{\mathrm{res}}^{\mathrm{Scat. Env.}}$ of its constitutive dipoles. If $f_{\mathrm{res}}^{\mathrm{Scat. Env.}}$ is orders of magnitude above the operating frequency, the environment essentially does not scatter the waves and is effectively transparent, such that we are effectively dealing with free space where only the LOS component exists (i.e., $K \rightarrow \infty$). 

By sweeping across the different values of $f_{\mathrm{res}}^{\mathrm{Scat. Env.}}$, we can therefore conveniently adjust the $K$-factor via a single parameter in our PhysFad model. Our focus here is on the easy tunability of $K$; of course, if the goal is to only simulate the case of $K \rightarrow \infty$ with a LOS link in free space (i.e., no fading), it is computationally more efficient to simply remove the dipoles that constitute the environment, as opposed to making them transparent. The most challenging part in sweeping all possible values of $K$
is to implement the Rayleigh condition (i.e., $K=0$) because even with a blocked LOS, some other short paths that are not affected by the stirring usually persist. We tackle this challenge by using a multitude of irregularly shaped mode-stirrers, as seen in the top left inset in Fig.~\ref{fig:RiceFigure}a. Of course, if the goal is to sweep through a range of $K$ values that does not include the Rayleigh condition ($K=0$), this can be implemented in a computationally more efficient manner by reducing the number of mode-stirrers and by optimizing their shape, size, and location; however, this is beyond the scope of this paper.

We summarize in Fig.~\ref{fig:RiceFigure} our results on tuning Rician fading from $K<-20~\mathrm{dB}$ to $K>50\ \mathrm{dB}$ by sweeping a single parameter: $f_{\mathrm{res}}^{\mathrm{Scat. Env.}}$. In Fig.~\ref{fig:RiceFigure}(a), we show that as $f_{\mathrm{res}}^{\mathrm{Scat. Env.}}$ is increased, $|\mu|$ is initially almost zero because the stirring process is very efficient and removes all static paths. As the transparency of the scattering environment begins to set in, the value of $|\mu|$ increases and then stabilizes once the scattering environment is already essentially transparent. In the same subfigure, we plot the dependence of $\sigma$ on $f_{\mathrm{res}}^{\mathrm{Scat. Env.}}$. $\sigma$ is initially constant and finite, but tends towards zero as the scattering environment's transparency sets in, and eventually only the LOS path remains significant. Consequently, the $K$-factor defined in Eq.~\eqref{eq:K} is initially constant at below $-20\ \mathrm{dB}$ for all $12$ considered channel coefficients in our $3\times4$ MIMO system. As the scattering environment's transparency sets in, $K$ increases towards infinity. For two iconic fading settings, we plot all obtained realizations of the channel coefficient in Fig.~\ref{fig:RiceFigure}(d,e). The probability density functions (PDFs) of the real and imaginary parts faithfully follow the expected Gaussian statistics. At the same time, the PDF of the magnitudes is well described through a Rician function.

In Fig.~\ref{fig:RiceFigure}(c), we plot the impact of sweeping the scattering environment's transparency on the effective rank $\langle R_{\mathrm{eff}}(\mathbf{H})\rangle$~\cite{roy2007effective} of the considered $3\times4$ channel matrix $\mathbf{H}$. Initially, the effective rank is high, albeit not at the maximum possible value of 3. Not reaching full rank is expected in random scattering environments; only by judiciously engineering the scattering environment with an RIS, it is possible to achieve full rank~\cite{del2019optimally,del2019optimized}. As the scattering environment's transparency sets in, $\langle R_{\mathrm{eff}}(\mathbf{H})\rangle$ decreases and eventually stabilizes very close to its minimum possible value of unity. Indeed, in free-space the channels are barely distinguishable from one another.

\begin{rem}
Even though the PDFs of the channel coefficient follow the desired Rician distributions in the above examples, this does \textit{not} imply that the above procedure could be replaced by generating a random matrix following the same distribution. The crucial difference is that each channel realization above is physically sound, whereas a random matrix knows nothing about space, causality, mesoscopic correlations, etc. (see Sec.~\ref{subsec:PhyFad}). In future work, we will implement other common fading models, like Weibull and Nakagami-$m$~\cite{hashemi1993indoor,nakagami1960m}, in a physically justified manner in PhysFad. Incidentally, a rich literature about statistical distributions measured in reverberation chambers already offers clear indications on how to implement Weibull in PhysFad~\cite{orjubin2006statistical}.
\end{rem}

\subsection{Algorithmic Summary of PhysFad}
\label{subsec:AlgSumm}

The above subsections detail the components combined by PhysFad into a channel model which translates a RIS-parametrized wireless environment into a physically-compliant end-to-end channel matrix representation with controllable level of fading. PhysFad represents all entities affecting the wireless communication channel as discrete dipoles. 
Below we summarize the steps required to generate wireless channel realizations using PhysFad:

\begin{enumerate}
  \item Identify the desired wireless scenario by representing it as a 2D horizontal slice: 
  
   \begin{enumerate}
         \item Specify the number and location of transceiving antennas.
         \item Define the geometry of the scattering environment (e.g., enclosure, obstacles, etc.).
         \item Discretize continuous surfaces of the scattering environment by representing them as dipole fences.
         \item Specify the number and location of the RIS elements.
  \end{enumerate}
  
  \item Fine-tune each dipole's parameters ($\chi, f_{\textrm{res}}, \Gamma^L$) to capture specific antenna characteristics, RIS designs, and/or environmental properties (e.g., wall reflectivity). 
  
  \item Define the available programmable states of the RIS elements (i.e., 1-bit (binary), multi-bit, or continuous) by assigning suitable options for the resonance frequency $f_{\textrm{res}}^{\textrm{RIS}}$.
  
  \item Input the desired RIS configuration and obtain the corresponding end-to-end channel matrix using the procedure illustrated in Fig.~\ref{fig:MatrixFigure}.
  
  \item To obtain multiple realizations of a fading channel:
    \begin{enumerate}
         \item Define what processes cause the fading (e.g., moving scatterers). For Rician fading, define a multitude of scatterers as in Fig.~\ref{fig:RiceFigure} and choose $f_{\mathrm{res}}$ for all dipoles constituting the scattering environment according to the desired $K$ value (see Fig.~\ref{fig:RiceFigure}(b)).
         \item Convert this information into PhysFad input parameters following the above steps.
  \end{enumerate} 
\end{enumerate}

\section{Time-Domain Representation}
\label{sec:TDRep}
An important feature of PhysFad is that it naturally incorporates a notion of space and causality which is essential to work with the channel impulse response (CIR) in the time domain. Causality is a universal principle that must be satisfied in any real system: the output of a system cannot temporally precede the input. Thereby, causality is intimately linked to a notion of space that is completely absent in statistical channel models (see Sec.~\ref{subsec:PhyFad}). 

PhysFad is formulated in the frequency domain, but because it inherently includes a notion of space and causality, it is sufficient to perform an inverse Fourier transform of the channel's spectrum $H(f)$ to obtain the corresponding physics-compliant CIR $h(t)$. 
Technically, care must be taken in the choice of window function applied to $H(f)$ to minimize windowing artefacts like sidelobes. To compute the received time-domain signal, one can either convolute the emitted signal with the CIR $h(t)$, or, more efficiently, multiply the emitted signal's spectrum with the transmission spectrum $H(f)$ before performing the inverse Fourier transform.
\begin{figure*}[!t]
    \centering
    \includegraphics[width=\columnwidth]{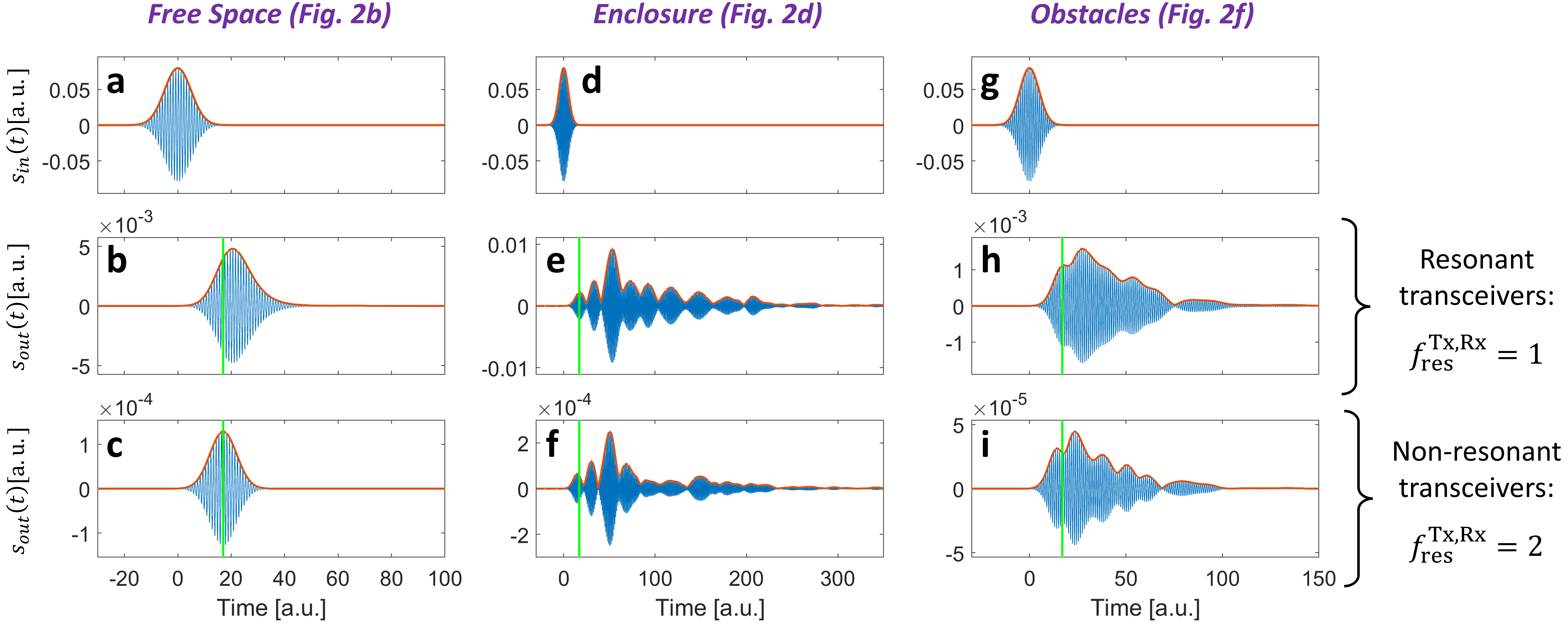}
    \caption{Time-domain analysis of the settings considered in Fig.~\ref{fig:BasicFieldPlots}. Upon transmission of a Gaussian pulse $s_{\text{in}}(t)$ (top row), we obtain the received signal $s_{\text{out}}(t)$ for the cases of resonant (middle row) and non-resonant (bottom row) transceivers. The vertical green line indicates the instant in time $t=D/c$, where $D$ is the transceiver separation. The horizontal axes are different in each column of the figure but the duration of the emitted pulses and the location of the vertical green line are the same in all cases.}
    \label{fig:TimeDomain}
\end{figure*}

We provide example CIRs for three illustrative settings in Fig.~\ref{fig:TimeDomain} to demonstrate the time-domain capabilities of PhysFad. First, we consider the transmission of a Gaussian pulse between two transceivers in free space. Naively, if the pulse is emitted at time $t=0$, one may expect the pulse to arrive at the receiver exactly at time $t=D/c$, where $D$ is the separation between the transmitting and receiving dipoles. Specifically, the maximum of the pulse should arrive at that time, but due to the finite bandwidth the signal rises and falls before and after this time, respectively. This naive assumption is exactly verified in Fig.~\ref{fig:TimeDomain}c for the case of two transceivers that are not resonant within the considered frequency interval. However, in Fig.~\ref{fig:TimeDomain}b, where we consider two resonant transceivers, the pulse arrives later than ``expected'' (and is slightly distorted). This deviation from the naive picture is in fact due to the well-understood interaction of pulses with resonators that results in pulse delays (see, e.g., Ref.~\cite{asano2016anomalous}. The fact that PhysFad faithfully captures such subtleties evidences once again how deeply routed it is in wave physics. We also reconsider the rich scattering scenarios in an enclosure and in a collection of obstacles. Therein, the CIR is seen to be lengthy due to severe multipath. Moreover, the CIR maximum is not associated with the shortest path but occurs at a later time at which many paths happen to interfere constructively. Finally, a difference in CIR shape related to the transceiver nature (resonant or not) is again apparent.

\section{Case Study: RIS-Enabled Over-the-Air Equalization}
\label{sec:CaseStudy}
In this section, we present a case study on RIS-enabled over-the-air channel equalization that is ideally suited to highlight the unique capabilities of PhysFad in capturing the \textit{time-domain} aspects of \textit{RIS-parametrized fading} environments. 
Shaping the CIR in a multipath environment to make it pulse-like (as if communication takes place in free space) is a form of over-the-air equalization that is relevant to scenarios with limited (de)modulation capabilities, such as in the Internet-of-Things or wireless networks on chips (WNoCs). If one operates with simple On-Off-Keying (OOK) schemes, a lengthy CIR due to multipath directly results in inter-symbol interference, unless one reduces the symbol rate.\footnote{While over-the-air equalization maximizes the information transfer rate with simple modulation schemes like OOK, it does \textit{not} necessarily maximize the modulation-scheme-independent channel capacity~\cite{alexandg_2021}.} The ability to impose pulse-like CIRs is hence highly desirable for OOK in fading environments.

Rigorous demonstrations of RIS-assisted over-the-air channel equalization were reported with experiments in the $2.5$ GHz regime~\cite{del2016spatiotemporal}, and recently based on full-wave simulations in the context of WNoCs~\cite{imani2021smart}. However, there is also a significant timely interest in studying these aspects based on channel models~\cite{zhang2021spatial,arslan2021over}. To date, such studies emulate fading through cascaded random matrices as outlined in Sec.~\ref{subsec:PhyFad}. But random matrices are ignorant of the spatial arrangement of transceivers, implying that such models inevitably do not obey causality which is a prerequisite for studying time-domain phenomena. This would be immediately obvious if the considered CIRs were plotted in the time domain. 

In the following, we report a rigorous study of RIS-enabled over-the-air equalization based on a channel model, namely PhysFad. The study's purpose is to demonstrate how PhysFad allows us to evaluate algorithms for challenging RIS-empowered communication problems involving fading, as well as that the achieved results are in line with wave physics and previous experimental results. To start, it is important to understand that the LOS path can never be altered by the RIS, because it does not interact with the RIS. Therefore, in aiming at a pulse-like CIR, either all NLOS taps must be suppressed through destructive interferences, or one NLOS tap must be substantially enhanced through constructive interferences so that the persisting LOS and other NLOS taps become negligible. The ability of the RIS to control NLOS taps depends on: \textit{i}) the amount of RIS elements (and their properties); and \textit{ii}) the amount of reverberation. The former is obvious, the latter is also intuitive: the more reverberation there is, the more often any given path is likely to interact with the RIS, and hence to be controllable. Incidentally, this dwell-time-enhanced path sensitivity is the basis of the recently reported deeply sub-wavelength localization~\cite{del2021deeply}.

We consider two qualitatively different regimes. First, we pursue the strategy of imposing \textit{one dominant NLOS channel tap} inside an irregularly shaped scattering enclosure with \textit{large amount of reverberation}; second, we reduce the amount of reverberation by adding loss to the dipoles constituting the scattering environment, and explore the alternative strategy of \textit{cancelling all NLOS taps} through destructive interference in this setting with \textit{lower amount of reverberation}. In both cases, we define our cost function as the ratio of signal intensity in the desired tap  of the CIR to the total signal intensity:
\begin{equation}
\mathcal{C} = \frac{\int_{t_0-\Delta t/2}^{t_0+\Delta t/2} h^2(t) \mathrm{d}t}{\int_0^\infty h^2(t) \mathrm{d}t},
\label{eq:CF}
\end{equation}
where $t_0$ is the peak time and $\Delta t$ is the width of the CIR tap that we desire to make the only significant tap. We use the simple iterative Algorithm~\ref{alg:over-the-air-equalization} to optimize the RIS configuration such that it maximizes $\mathcal{C}$, where PhysFad is used in each iteration to generate the CIRs needed to compute $\mathcal{C}$ via Eq.~\eqref{eq:CF}.
To be clear, our contribution is not  Algorithm~\ref{alg:over-the-air-equalization} itself, which has already been used in Refs.~\cite{del2016spatiotemporal,imani2021smart}, but its implementation with our physically justified channel model, namely PhysFad.
Algorithm~\ref{alg:over-the-air-equalization} chooses the best out of $50$ random RIS configurations and then tests element-by-element if flipping its state increases $\mathcal{C}$. Multiple loops over all RIS elements are needed because mesoscopic long-range correlations mean that the optimal configuration of a given RIS element depends on the configuration of \textit{all} other RIS elements.
For each setting, we study distributed binary RIS with different numbers of elements.

\begin{algorithm}[!t]
	\caption{Binary RIS Optimization for Over-the-Air Equalization}
	\label{alg:over-the-air-equalization}
	Evaluate $\{\mathcal{C}_i\}$ for $50$ random RIS configurations $\{C_{i}^0\}$.\\
	Select configuration $C_{\rm curr}$ corresponding to $\mathcal{C}_{\rm curr} = \mathrm{min}_i(\{\mathcal{C}_i\})$.\\
	\For{$i=1,2,\ldots,5N_{\rm RIS}$}{
	    Define $C_{\rm temp}$ as $C_{\rm curr}$ but with configuration of $\mathrm{mod}(i,N_{\rm RIS})$th RIS element flipped.\\
	    Evaluate $\mathcal{C}_{\rm temp}$.\\
	    	        \If{$\mathcal{C}_{\rm temp} > \mathcal{C}_{\rm curr}$}{
	            Redefine $C_{\rm curr}$ as $C_{\rm temp}$ and $\mathcal{C}_{\rm curr}$ as $\mathcal{C}_{\rm temp}$.
	        }
	}
	\KwOut{Optimized RIS configuration $C_{\rm curr}$.}
	\label{alg:opti}
\end{algorithm}
\begin{figure*}[!t]
    \centering
    \includegraphics[width=\columnwidth]{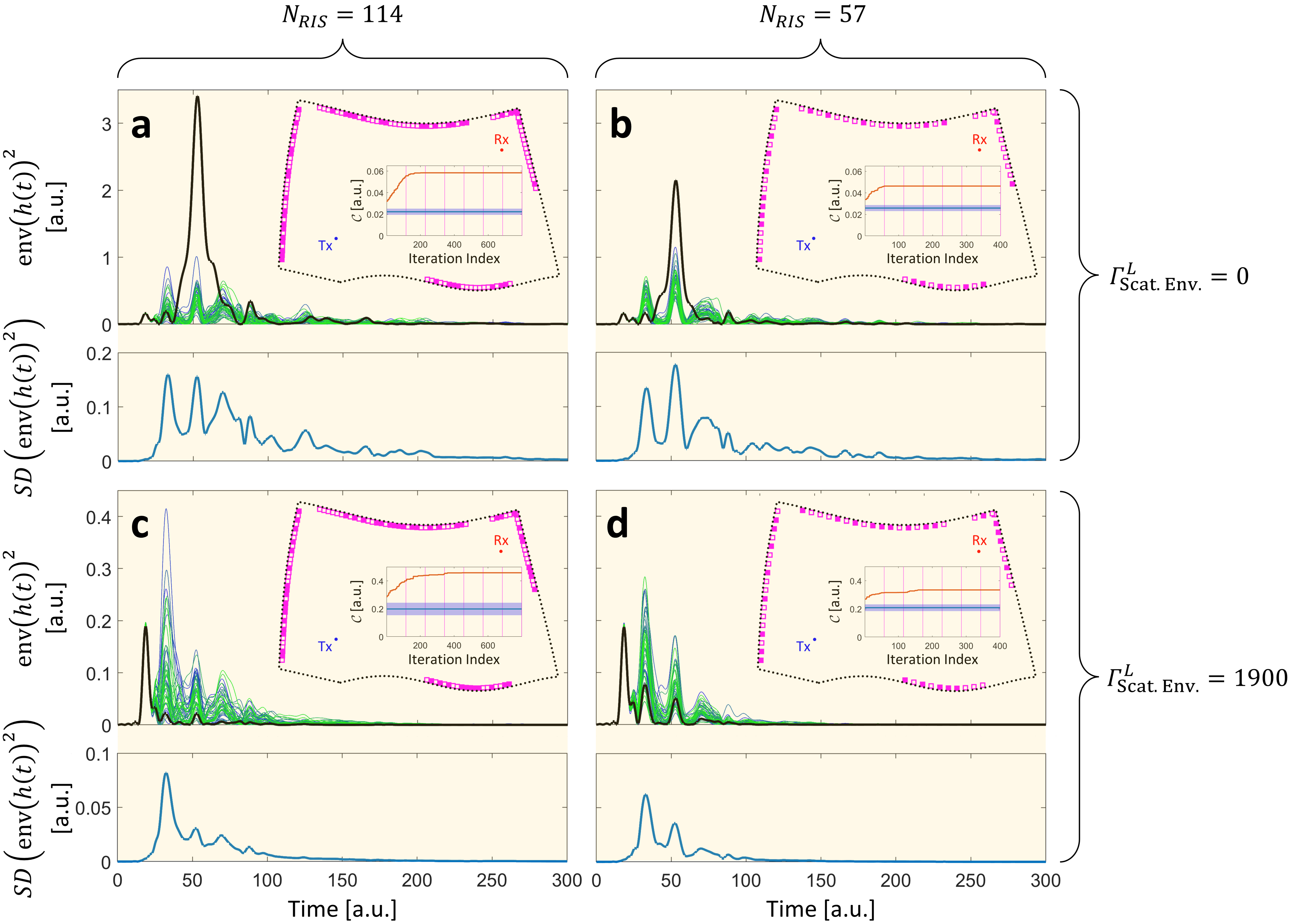}
    \caption{Over-the-air channel equalization enabled by RISs in PhysFad based on Algorithm~1. We show four representative scenarios for environments with high (a,b) and low (c,d) amounts of reverberation, in each case for $N_{\rm RIS}=114$ (a,c) and $N_{\rm RIS}=57$ (b,d). For each example, we show the CIR intensities corresponding to $50$ random RIS configurations (color-coded) and to the optimized RIS configuration (black), as well as the standard deviation over the former (bottom). The inset shows the considered setup including the optimized RIS configuration (filled symbol represent RIS elements configured to be resonant at the operating frequency). Additionally, an inset indicating the optimization dynamics is provided: the blue line and area represent average and standard deviation over the 50 random RIS configurations, and the red line the subsequent iterative optimization.
    }
    \label{fig:Opti}
\end{figure*}

Representative results from this case study are synthesized in Fig.~\ref{fig:Opti}. For the scenario with a high amount of reverberation, Fig.~\ref{fig:Opti}(a) shows that it is possible to optimize the RIS such that one NLOS tap clearly dominates all other taps (LOS and NLOS) by at least one order of magnitude. Thereby, with appropriate synchronization, it is possible to communicate via OOK modulation without inter-symbol interference, despite operating in a strongly multipath environment. Moreover, the desired received signal strength is substantially enhanced. As expected, using fewer RIS elements deteriorates the over-the-air equalization performance, as seen in Fig.~\ref{fig:Opti}(b). For the scenario with a lower amount of reverberation, Fig.~\ref{fig:Opti}(c) shows that it is possible to optimize the RIS such that only the LOS tap remains significant. Moreover, we note that in most cases the optimization has not converged after $N_{\rm RIS}$ iterations, which is clear evidence of long-range correlations between the optimal configuration of different RIS elements. The impact of the RIS on the CIR can hence not be approximated in a linear fashion, justifying the need for the iterative trial-and-error Algorithm~\ref{alg:over-the-air-equalization}. We also empirically observe that running Algorithm~\ref{alg:over-the-air-equalization} multiple times yields different outcomes of similar quality.

\section{Open Source Reusable Code}
\label{sec:Software}
Our goal is to encourage wireless communication practitioners to develop novel algorithms for RIS-parametrized fading channels based on channel models that are faithful to wave physics. To that end, we do not limit our paper to describing a suitable channel model, but we take active steps toward helping the wireless community to deploy PhysFad in their future work. Indeed, the works cited in Sec.~\ref{subsec:Free} provide mathematical expressions related to RIS-parametrized channel models in free space  \textit{without fading}, but the barrier toward deployment is still high if community members have to first understand such papers in every mathematical detail and then write their own codes to implement the models. In contrast, we share in Ref.~\cite{GitHub} our source code to determine the end-to-end channel matrix in an exemplary rich-scattering RIS-parametrized environment. Our open-code approach endows the community with an easy-to-use tool to integrate PhysFad into algorithmic design processes without a need for extensive coding.

\section{Conclusions and Outlook}
\label{sec:Conc}
PhysFad provides a physically justified channel model for RIS-parametrized wireless environments \textit{with adjustable fading}. In this paper, we have detailed its principles, highlighted some of its features related to adjustable fading, as well as causality and the time-domain representation. Moreover, we provided a prototypical case-study demonstration of using PhysFad to evaluate a signal processing technique in the context of RIS-enabled over-the-air channel equalization. We also share associated codes openly with the wireless community to facilitate the use of PhysFad.

Looking forward, on the one hand, we envision that PhysFad in its current form will serve as basis for many RIS-related generic algorithmic explorations. On the other hand, we expect that \textit{i}) PhysFad can be upgraded to a dyadic 3D version, and that \textit{ii}) detailed models, based on the coupled-dipole formalism, of specific antenna and RIS designs will emerge, enabling simulations of specific wireless systems that are faithful to the underlying wave physics. Moreover, we foresee that PhysFad becomes a valuable tool in backscatter communications which, being based on impedance-modulated antennas such as RFID tags, are conceptually closely related to RISs~\cite{zhao2020metasurface,f2020perfect,vardakis2021intelligently}.
Beyond wireless communication simulations, PhysFad will also serve in the areas of mesoscopic wave physics and (extreme) wave scattering.

\section{Acknowledgment}
P. del Hougne thanks S. M. Anlage, T. M. Antonsen, M. Davy, R. Fleury, E. Ott, and Z. Peng for stimulating discussions.

\bibliographystyle{IEEEtran}


\end{document}